\documentclass{edp-jp4}
\usepackage{graphicx}
\usepackage{bm}

\begin{document}

\title{Dynamics of 8CB confined into porous silicon probed by incoherent neutron backscattering experiments.} 

%

\author{R\'egis Gu\'egan}\address{Groupe Mati\`ere Condens\'ee et Mat\'eriaux, UMR-CNRS 6626, Universit\'e de Rennes 1, 35042 Rennes, France}
\author{Ronan Lefort}\sameaddress{1}
\author{Wilfried B\'eziel}\sameaddress{1}
\author{Denis Morineau}\sameaddress{1}
\author{Mohammed Guendouz}\address{Laboratoire d'Optronique, FOTON, UMR-CNRS 6082, 22302 Lannion, France}
\author{Bernhard Frick}\address{Institut Laue Langevin, B.P. 156, 38042 Grenoble, France}

\thanks{corresponding author : denis.morineau@univ-rennes1.fr}

\maketitle
\begin{abstract}

Confinement in the nanochannels of porous silicon strongly affects the phase behavior of the archetype liquid-crystal 4-n-octyl-4-cyanobiphenyl (8CB). A very striking phenomenon is the development of a short-range smectic order, which occurs on a very broad temperature range. It suggests in this case that quenched disorder effects add to usual finite size and surface interaction effects. We have monitored the temperature variation of the molecular dynamics of the confined fluid by incoherent quasielastic neutron scattering. A strongly reduced mobility is observed at the highest temperatures in the liquid phase, which suggests that the interfacial molecular dynamics is strongly hindered. A continuously increasing slowdown appears on cooling together with a progressive growth of the static correlation length.

 \end{abstract}
%
\section{Introduction}

The understanding of the special behavior of molecular fluids confined in nanoscopic cavities has gained an increasing interest because of the number of fundamental questions they raise. The scientific activity in this field has been recently amplified by the emergence of nanosciences and nanotechnologies.
Because it reaches nanometer sizes, mesoporous confinement introduces new properties to the fluid, which are not simply deduced from the bulk ones. Mostly addressed issues concern the phase behavior and especially the freezing/melting transitions\cite{Christenson-2001}. Another very active topic concerns the molecular dynamics of confined fluids and its relation with the glass transition \cite{Confit-2003}.
 Two major effects are usually invoked in mesoporous confinement. The first is cut-off or finite size effect. It implies that neither static nor dynamical correlation length can grow larger than the maximum pore size. The second one is surface effect, which is introduced by the large surface-to-volume ratio encountered in system of nanometer-size.     
A better understanding of the physical properties of confined simple fluids has been achieved by considering these two effects. An improved supercooling is achieved in small volumes, which reduces the probability of nucleation. A simple consequence of finite size effects is the observed restriction of translational orders, which results in a broadening of the crystalline Bragg peaks and excluded volume effects for the structure factor of liquids\cite{Morineau-JCP-2003}. Surface interaction adds to the volume energy of the system, which yields to the Gibbs-Thomson theory for the melting point variation in the limit of a macroscopic thermodynamic approach. Considering the importance of fluid-wall interactions with respect to fluid-fluid ones allows one to envisage a large amorphous frozen interfacial component in nanoconfined crystals or, for special substrates, the occurrence of ordered contact layers\cite{Gubbins-JCP-2003}. Additional effects, such as pore dimensionality, should not be ignored in some special cases for instance for long alkanes solid phases or for marginally anisotropic confined liquid alcohols\cite{Huber-EPL-2004}\cite{Guegan-CP-2005}.    
The molecular dynamics of confined liquids is certainly more complex than their thermodynamics is. Nonetheless, there is growing number of evidences that interfacial interactions effects markedly dominate the structural relaxation of globular glass-forming liquids. The novel dynamical boundary condition introduced by the pore surface generally reduces the molecular dynamics. The large increase of the local structural relaxation time is transmitted to the inner fluid through dynamical molecular correlations. This leads to the globally slower and inhomogeneous dynamics of the confined fluid. A crucial issue is the nature of the correlation length associated with this surface-induced spatially heterogeneous dynamics. It is commonly accepted that it is indeed the cooperativity length of the relaxation dynamics of the confined liquid. Variable diameter confinement experiments have been thought to test a possible cooperativity length associated to the glassy dynamics of supercooled liquids. This supposes of course that the thermodynamical state of the bulk liquid is unmodified in confinement, which does not seem to be the usual case\cite{Morineau-JCP-2002}. Eventually, finite size has to be considered as the other dominant effect that may affect the dynamics of supercooled liquids.  Most likely, the only modest increase of the cooperativity length in supercooled liquids requires very small confinement diameter for finite size effects to be effective, which precludes surface effects and the thermodynamical state of the confined phase to be fully controlled. 
Molecular fluids relevant for technological applications based on nanoconfinement (fluidics, optronics, biotechnology…) are usually complex systems including soft matter and bio-solutions. They display additional features related to molecular self-assembling, various phase transitions and multiple relaxation processes. They bring about new challenging questions for fundamental research on nanoconfined molecular phases. Additional effects are to be considered, such as quenched disorder and low dimensionality. They are of special interest when orientational correlations or continuous phase transitions issues are concerned \cite{Aliev-JNCS-2001}, \cite{Bellini-Science-01}. 
Beyond this further level of molecular complexity, confined soft matter has provided model systems to address some fundamental questions arising from statistical mechanics. In this field, smectic liquid crystals (LC) confined into disordered porous materials are unique systems to investigate the effect of quenched disorder introduced by the porous matrix on a continuous symmetry breaking transition. Huge growth of dynamical and static correlation lengths are usually encountered in mesogenic liquids. They provide a way to study the behavior of a fluid during a gradual temperature-controlled transmission of the surface boundary conditions imposed by confinement to the entire pore volume. In the present paper, we address the case of a mesogenic fluid, which develops a short-range order when confined in nanochannels. Consequences on the molecular dynamics are analyzed on the basis of incoherent quasielastic neutron scattering experiments.

\section{Structure and phase behavior}

We have recently described some aspects related to the structure and the phase behavior of a smectic liquid crystal 4-n-octyl-4-cyanobiphenyl (8CB) confined in porous silicon (p-Si)\cite{Guegan-PRE-2006}. The columnar form of p-Si provides a model (1D)-porous geometry in terms of macroscopically aligned channels of mean pore diameter of 300\AA{} and length 30$\mu$m as shown in Fig. 1.
This peculiar porous wafer geometry  gives an unique opportunity to study how complex fluids with marked orientational correlations behave in a one-dimensional confinement. The anisotropy of the structural and dynamical parameters of the confined fluid can be measured without powder average effects by a simple alignment of the porous silicon wafers with respect to the incident neutron beam (cf. Fig.~\ref{FIG-cell}). 

Although the main axis of the nano-channels of p-Si are straight and aligned to each other up to macroscopic distances, their inner surface exhibits a large corrugation at the nanometer scale \cite{Lehmann-MS-00}. The geometric restriction imposed by p-Si to the confined phase therefore differs from other 1-D mesoporous geometries (anopore, nuclepore). It has been shown to introduce quenched disorder effects, which affect the capillary condensation \cite{Knorr-PRL-04} and the smectic transitions \cite{Guegan-PRE-2006}. Such features had been most commonly reported for random porous silica materials but not for (1D)-channels so far \cite{Bellini-Science-01},\cite{Leheny-PRE-03}.

\begin{figure}[htp]
\label{FIG-meb}
\includegraphics[width=0.5\columnwidth]{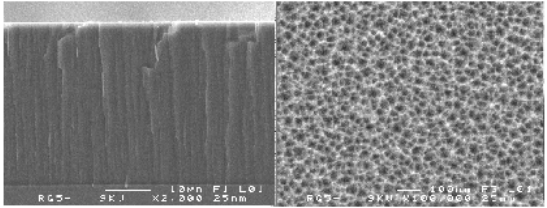}
\caption{Scanning electron micrographs of the porous silicon film. (left) side view at low magnification showing the 30 $\mu$m thick porous layer attached to the silicon substrate. (right) top view at higher magnification.}
\end{figure}

Neutron diffraction experiments have been performed in order to analyze the structure of the confined fluid with the cold-source double-axis diffractometer G6.1 at the LLB. The phase sequence of the pure component is isotropic (I), nematic (N) smectic A (SmA) and crystal (K) with equilibrium transition temperatures being respectively $T_{NI}=313 K$, $T_{NA}=306.7 K$ and $T_{KA}=294 K$. A first effect of confinement is the preferential macroscopic alignment of both orientational and translational orders along the direction of the pores. This feature related to anchoring effects has been discussed for other 1D-pores, such as porous alumina (anopores) and track-etched membranes (nuclepores). More surprising is the extreme alteration of the phase diagram of 8CB in p-Si. Crystallization is strongly depressed and the smectic transition disappears. It is replaced by a reversible gradual increase of a short-range translational local order that develops on an extremely broad temperature range of about 50 K. This local smectic order is characterized by a single broad diffraction peak at $q=0.2 $ \AA$^{-1}$, which corresponds to the location of the smectic Bragg peak in the bulk \cite{Ocko-ZPB-86}. The analysis of the shape of this broad peak requires a two components formula, which has been theoretically motivated by random fields theories. The origin of this behavior can be conceived in the frame of quenched disorder effects due to the disrupted interactions between the fluid and the inner pore surface of p-Si,\cite{Guegan-PRE-2006}. It corroborates the fact that a transition breaking a continuous symmetry is unstable toward arbitrarily small quenched random fields \cite{Radzihovsky-PRE-99}. In the case of a smectic LC, the fluid develops a static correlation length that saturates to short-range distances at low temperatures. This effect has been essentially studied for random porous materials such as aerogels and aerosils-dispersions \cite{Bellini-Science-01}\cite{Leheny-PRE-03}. Note that in the case of LC confined in aerosils-dispersions, the smectic correlation length does not vary significantly below the pseudo-transition temperature, but the order parameter keeps increasing at low temperatures as for aerogels and p-Si.\\

\begin{figure}[htp]
\label{FIG-cell}
\includegraphics[width=0.3\columnwidth]{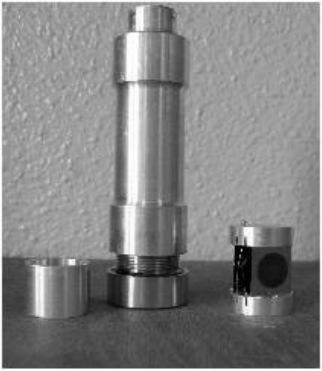}
\caption{Photograph of the sample container designed for neutron scattering experiments, in order to control the orientation of the eight aligned porous silicon wafers with respect to the neutron beam.}
\end{figure}

\begin{figure}[htp]
\includegraphics[width=0.5\hsize]{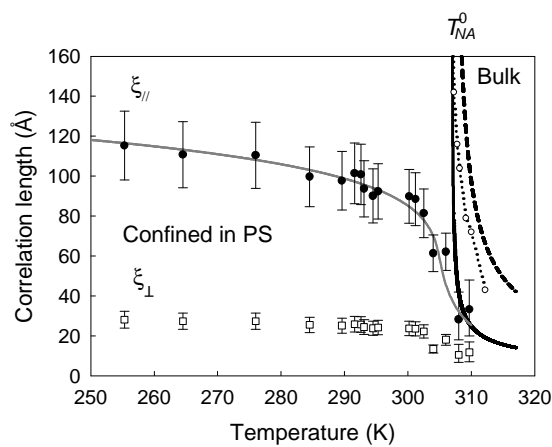}
\caption{\label{FIG-ksi} Temperature variation of the smectic correlation lengths $\xi_{\|}$ (filled circles) and $\xi_{\bot}$ (open squares) for 8CB confined in porous silicon \cite{Guegan-PRE-2006}. The gray solid line is a guide for the eyes. The same quantities for bulk 8CB are plotted in dashed and solid lines respectively \cite{Ocko-ZPB-86}. The smectic correlation lengths $\xi_{\|}$ for 8CB confined in porous alumina (anopore) is also displayed for comparison (open circles with dotted line).}
\end{figure}

Fig.~\ref{FIG-ksi} shows the temperature variation of the correlation length of bulk 8CB and confined in p-Si. When approaching the smectic transition from above, the bulk correlation length increases due to pretransitional fluctuations and critically diverges at ($T^0_{NA}=306.7 K$). A similar scenario is observed for 8CB confined in alumina anopores of diameter 200\AA {} with a smooth pore surface (open circles). The transition is only slightly depressed due to surface interactions (Gibbs-Thomson effects) and the true divergence is aborted by finite size effects. In addition to usual surface and finite size effects, quenched disorder introduced by p-Si leads to a progressive and moderate increase of the correlation length from molecular size at $T_{NA}$ to about 150 \AA {} at 250 K. This temperature variation and the value of the saturating correlation length is expressed theoretically in terms of a competition between the strength of random field disorder and the elasticity of the confined phase \cite{Radzihovsky-PRE-99}.

\section{Molecular dynamics}

\begin{figure} [htb]
\includegraphics[width=0.5\hsize]{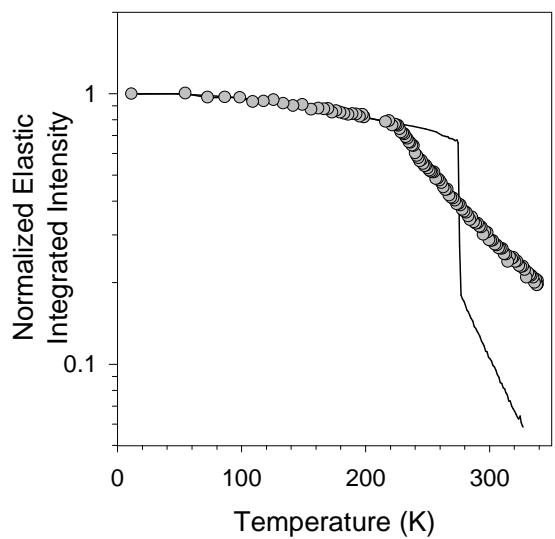}
\caption{\label{FIG-elastic} Elastic scattering of 8CB confined in porous Si (circles) or bulk (solid line). The intensity is corrected for empty sample contribution, integrated from 0.43 to 1.89 \AA$^{-1}$ and normalised at 10K.}
\end{figure}

The progressive increase of a locally preferred order, which satisfies the boundary conditions imposed to the confined fluid raises questions about the variation of the molecular relaxation dynamics. The individual molecular dynamics has been probed by incoherent quasi-elastic neutron scattering using the high-resolution backscattering spectrometer IN16 (BS) at ILL. Elastic window measurements (FWHM of 0.9 $\mu$eV) have been performed along variable temperature scans with Doppler at arrest. Fig.~\ref{FIG-elastic} displays the elastic intensity measured on cooling for bulk 8CB (solid line) and confined in p-Si (circles). Scans of the empty cell and the empty Si wafers have been subtracted for background correction and the spectra integrated on a q-range from 0.43 to 1.89 \AA$^{-1}$.  
In the bulk above 300 K, the elastic intensity is very small (about 0.05), which is typical for a liquid where relaxation processes correspond to very broad quasi-elastic lines. On cooling from the bulk liquid, the elastic intensity firstly increases progressively down to about 275 K where it suddenly jumps to larger values.

Then it keeps increasing moderately down to the lower temperature measured. The sharp discontinuity observed on cooling reflects the phase transition from mesomorphic phases to a crystalline phase. In the high temperature region, the fast molecular relaxation processes probed by these techniques progressively slow down as the temperature decreases. It should be noted that this individual molecular dynamics is not strongly affected by the liquid-crystal ordering since no signature of the isotropic-nematic nor the nematic-smectic transition are observed. This is at variance with the crystallization phenomenon, which is detected as a sharp discontinuity.  Most relaxation processes probed by this technique are actually frozen in the crystalline state. The small temperature dependence of the elastic component in the crystal phase reflects remaining  degrees of freedom of the crystal (methyl libration and Debye-Waller factor, comprising inter- and intramolecular vibrational modes). 

\begin{figure}[htb]
\includegraphics[width=0.5\hsize]{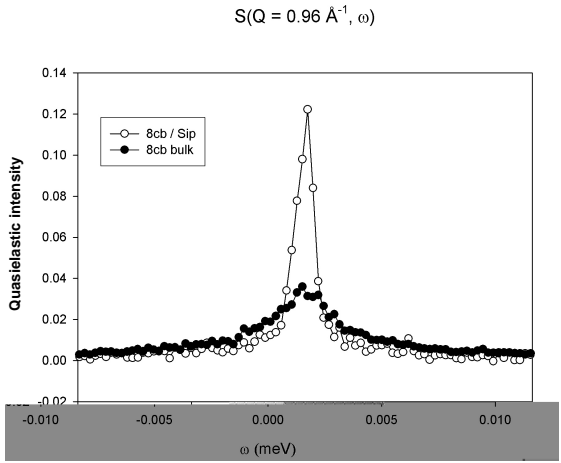}
\caption{\label{FIG-quasi} Quasielastic spectra measured at $q=0.96 $ \AA$^{-1}$ and 296 K for bulk 8CB (black circles) and 8CB confined into porous silicon (white circles).}
\end{figure}

The elastic intensity of confined 8CB presents a very different behavior. Indeed, its value is always large, even in the liquid phase, being equal to 0.2 at 315 K. In addition, it progressively increases on decreasing temperature down to 220 K, where it bends as it reaches the bulk crystalline branch. The absence of any discontinuity supports the large depression of the crystallization process. This dynamical behavior is strongly dominated by the surface effects and the unusual phase behavior of 8CB confined in p-Si. The unexpected high value of the elastic intensity well above the clarification point is most probably a consequence of the strong reduction of the molecular mobility introduced by the interfacial boundary conditions\cite{Schoenhals-CPL-99}. This shows up in the quasi-elastic spectra recorded at 295 K in Fig.~\ref{FIG-quasi}. A large elastic component is observed in the spectrum recorded in confined geometry whereas the bulk one is fully quasi-elastic. In addition, the temperature variation reflects the different phase behaviour of bulk and confined 8CB. We have recently shown that 8CB confined in p-Si progressively transforms to a short-range-ordered smectic phase down to 250 K with no crystallization\cite{Guegan-PRE-2006}. The temperature variation of the elastic intensity clearly demonstrates that this unusual phase behaviour goes with a progressive slowing down of the individual molecular dynamics.

\section{Conclusion}
8CB confined in p-Si shows an unusual phase behavior. It develops a short range smectic order on a large temperature range (about 50 K), which has been attributed to quenched disorder effects. 

Two main consequences of confinement on the dynamical behavior of 8CB are observed within the high-energy resolution provided by BS experiments. The first one is a global slowing down of the relaxation dynamics of the confined fluid phases as compared to the bulk one. This is partly attributable to the surface-induced dynamical boundary conditions. The second consequence is a continuous slowing down on cooling, which compares with the expansion of short-range structural correlations in the confined 8CB.






\begin{thebibliography}{16}

\bibitem{Christenson-2001}
H.~K. Christenson,
\newblock {\em J.\ Phys.:\ Condens.\ Matter} \textbf{13},R95 (2001).

\bibitem{Confit-2003}
for a review~see :
\newblock {\em Eur.\ Phys.:\ J.\ E} \textbf{12},1 (2003).

\bibitem{Morineau-JCP-2003}
D.~Morineau and C.~Alba-Simionesco,
\newblock {\em J.\ Chem.\ Phys.} \textbf{118}, 9389 (2003).


\bibitem{Gubbins-JCP-2003}
R.~Radhakrishnan, K.~Gubbins, and M.~Sliwinska-Bartkowiak,
\newblock {\em J.\ Chem.\ Phys.} \textbf{116}, 1147 (2003).

\bibitem{Huber-EPL-2004}
P.~Huber, D.~Wallacher, J.~Albers, and K.~Knorr,
\newblock {\em Eur.\ Phys.\ Lett.} \textbf{65},351 (2003).

\bibitem{Guegan-CP-2005}
R.~Gu\'egan, D.~Morineau, and C.~Alba-Simionesco,
\newblock {\em Chem.\ Phys.} \textbf{317}, 236 (2005).


\bibitem{Morineau-JCP-2002}
D.~Morineau, Y.~Xia, and C.~Alba-Simionesco,
\newblock {\em J. \ Chem.\ Phys.} \textbf{117}, 8966 (2002).

\bibitem{Aliev-JNCS-2001}
F.~M. Aliev, M.~Rivera Bengoechea, C.~Y. Gao, H.~D. Cochran and S. ~Dai,
\newblock {\em J.\ Non-Cryst.\ Solids} \textbf{351}, 2690 (2003).


\bibitem{Bellini-Science-01}
T.~Bellini, L.~Radzihovsky, J.~Toner, and N.~A. Clark,
\newblock {\em Science} \textbf{294}, 1074 (2001).



\bibitem{Guegan-PRE-2006}
R.~Gu\'egan, D.~Morineau, C.~Loverdo, W.~Beziel, and M.~Guendouz,
\newblock {\em Phys. \ Rev.\ E} \textbf{73}, 011707 (2006).


\bibitem{Lehmann-MS-00}
V.~Lehmann, R.~Stengl, and A.~Luigart,
\newblock {\em Mater.\ Sci.\ Eng.} \textbf{B69-70}, 11 (2000).


\bibitem{Knorr-PRL-04}

D.~Wallacher, N.~K{\"u}nzner, D.~Kovalev, N.~Knorr, and K.~Knorr,
  {\em Phys.\ Rev.\ Lett.} \textbf{92},195704 (2004).


\bibitem{Leheny-PRE-03}
R.~L. Leheny, S.~Park, R.~J. Birgeneau, J.-L. Gallani, C.~W. Garland, and G.~S.
  Iannacchione,
\newblock {\em Phys.\ Rev.\ E} \textbf{67}, 011708 (2003).


\bibitem{Ocko-ZPB-86}
B.~M. Ocko, R.~J. Birgeneau, and J.~D. Litster,
\newblock {\em Z.\ Phys.\ B:\ Condens.\ Matter} \textbf{62}, 487 (1986).





\bibitem{Radzihovsky-PRE-99}
L.~Radzihovsky and J.~Toner,
\newblock {\em Phys.\ Rev.\ B} \textbf{60}, 206 (1999).

\bibitem{Schoenhals-CPL-99}
S.~Frunza, L.~Frunza, A.~Schoenhals, H.-L. Zubowa, H.~Kosslick, H.-E. Carius,
  and R.~Frick,
\newblock {\em Chem.\ Phys.\ Lett.} \textbf{307}, 167 (1999).

\end{thebibliography}

\end{document}